\pdfoutput=1
\documentclass[12pt]{iopart}

\usepackage{iopams}

\usepackage{epstopdf}
\usepackage{psfrag}
\usepackage{ps4pdf}

\usepackage{multirow}
\usepackage{paralist}
\usepackage{color}

\begin{document}

\newcommand{\D}{\mathrm{d}}

%\emph{SHORT COMMUNICATION}

\title{Effective measurement height of atom gravimeters}
\author{V D Nagornyi}
\address{Metromatix, Inc., 111B Baden Pl, Staten Island, NY 10306, USA}
\ead{vn2@member.ams.org}
\begin{abstract}
The effective measurement heights of the fountain and the release types of atom gravimeters have been determined using the description of the instruments as linear time-invariant (LTI) system. The distance from the initial position of the atoms to the locus of the effective measurement height is expressed as simple fraction of the total trajectory length. Leaving the measured gravity for reporting at the effective measurement height allows to eliminate the gradient from calculations and avoid up to 1 $\mu$Gal of its uncertainty added to the instrument error budget.
\end{abstract}
\vspace{1cm}
\section{Introduction}
Atom gravimeters, like other types of ballistic gravimeters, require the measurement result to be corrected for the vertical gravity gradient. The  gradient in the correction formulas serves as influence quantity, and, according to the GUM (JCGM 100:2008) \cite{GUM2008}, its uncertainty should be included in the instrument's overall uncertainty budget. For many measurement sites with imprecisely known gravity gradient, this uncertainty component can become a significant portion of the overall uncertainty, and in certain cases even dominate it. Absolute gravimeters based on the tracking of macroscopic falling objects can generally avoid this problem by referring the measurement result to the so called ``effective measurement height'', on which the measured gravity does not depend on its vertical gradient. The problem, however, can not be avoided in atom gravimeters, for which the effective measurement height is unknown. This communication determines the effective measurement height for both the``release'' and the ``fountain'' types of atom gravimeters and finds the uncertainty component introduced by referring the measured value to the atoms' initial position.
\section{Atom gravimeter as LTI system}
\begin{figure}[ht]
\centering
%\small
\input{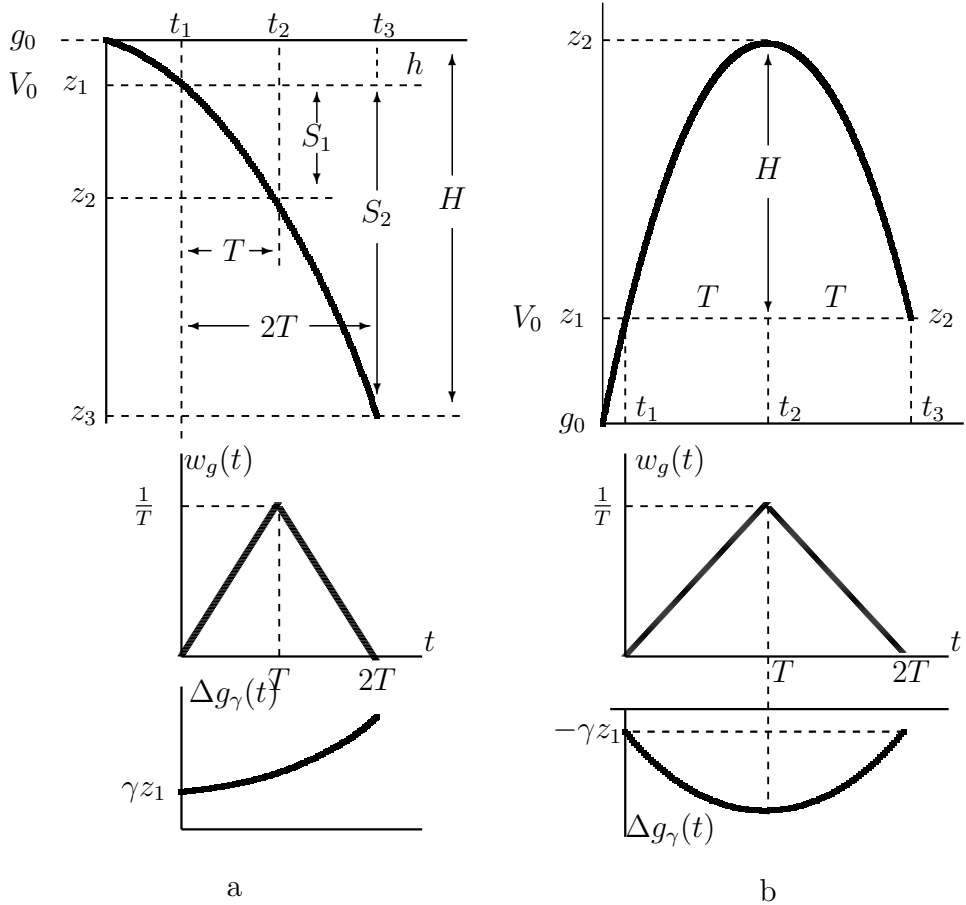}
  \caption[short title]
  {
  \quad\parbox[t]{10cm} {Trajectory model of atom gravimeters: a -- release type, b -- fountain type.
%      \begin{compactdesc}
%      \item[a: ]{release type,}
%      \item[b: ]{fountain type.}
%      \end{compactdesc}
  Also shown are weighting functions of the gravimeters by acceleration $w_g(t)$ and the disturbances due to the vertical gravity gradient $\Delta g_{\gamma}(t)$.
    }
  }
\label{fig_AG_SCHEMA}
\end{figure}

Atom gravimeters deduce the gravity acceleration from the phase change of the matter waves associated with free falling atoms. To measure the change, the group of atoms is first split and then recombined using a structure similar to the Mach-Zender interferometer. Given the phases $\phi_1$, $\phi_2$, $\phi_3$ of the matter wave at the moments $t_1$, $t_2$, $t_3$ separated by the time interval $T$, the measured gravity can be viewed as obtained from the formula \cite{peters2001}
\begin{equation}
\label{eq_3 phasis}
\overline g = (\phi_3 - 2\phi_2 + \phi_1)/k T^2,
\end{equation}
where $k$ is the matter wave number. This treatment is essentially equivalent \cite{peters1998} to the measurement of the positions $z_1$, $z_2$, $z_3$ of the falling atom separated by the time interval $T$ and calculating the gravity as
\begin{equation}
\label{eq_3 points}
\overline g = (z_3 - 2z_2 + z_1)/ T^2.
\end{equation}
Introducing the distances traveled by the atom as $S_1=z_2-z_1$, $S_2=z_3-z_1$, and corresponding time intervals as $T_1=T$ and $T_2=2T$, makes (\ref{eq_3 points}) equivalent to the 3-level measurement schema (fig.\ref{fig_AG_SCHEMA}) with the working formula
\begin{equation}
\label{eq_3 points}
\overline g = \left(\frac{S_2}{T_2}-\frac{S_1}{T_1} \right) \frac{2}{T_2 - T_1}.
\end{equation}
The above formula describes the gravimeter as linear time-invariant (LTI) system, so
the methods of LTI-analysis developed for macroscopic ballistic gravimeters \cite{zanimonskii1992, nagornyi1995, nagornyi2011} can be applied to atom gravimeters, at least for the analysis of disturbances common for both types of instruments.

\begin{figure}[ht]
\centering
%\small
\input{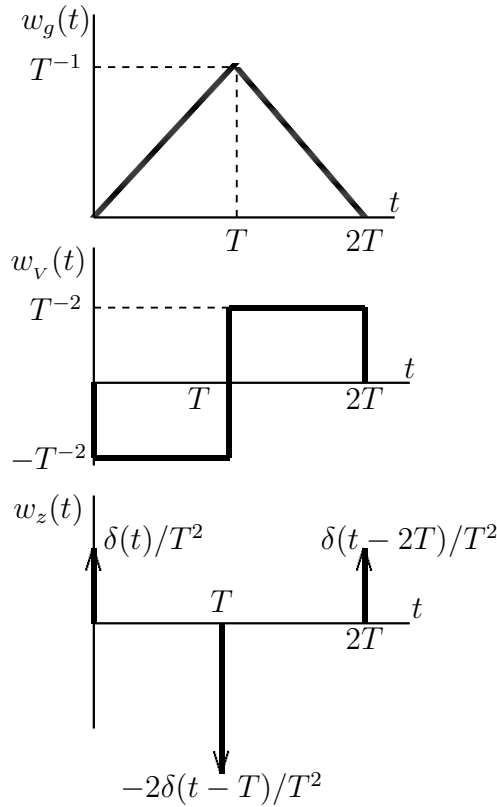}
  \caption[short title]
  {
  \quad\parbox[t]{6cm} {Weighting functions of atom agrvimeters (top to bottom): by acceleration, by velocity, by displacement.
    }
  }
\label{fig_AG_WFs}
\end{figure}

If the atom's acceleration changes during the measurement like
\begin{equation}
\label{eq_g_t}
g(t) = g_0 + \Delta g(t),
\end{equation}
the additional measured component caused by the term $\Delta g(t)$ can be found as
\begin{equation}
\label{eq_dg_meas}
\overline{\Delta g} = \int_0^{2T} \Delta g(t) w_g(t) \D t,
\end{equation}
where $w_g(t)$ is the weighting function of the gravimeter, which for the 3-level schema (\ref{eq_3 points}) is \cite{zanimonskii1992, nagornyi1995, nagornyi2011}
\begin{equation}
\label{eq_w_3_level}
w_g(t)=\cases
{
\frac{t}{T^2} & $0\le t \le T$,
\\
\frac{2}{T} - \frac{t}{T^2} & $T \le t \le 2T$ .
}
\end{equation}
If the disturbance is expressed in terms of changing velocity or coordinate, the additional measured component can still be found as in (\ref{eq_dg_meas}), but instead of $w_g(t)$, the weighting functions by velocity $w_{{}_V}(t)$ or coordinate $w_z(t)$ should be used (fig.~\ref{fig_AG_WFs}).
As
\begin{equation}
\label{eq_wz_wv_wg}
w_z(t) = -\frac{\D}{\D t}w_{{}_V}(t) =
-\frac{\D}{\D t}\left(-\frac{\D}{\D t} w_g(t)\right),
\end{equation}
the disturbance component can also be found as
\begin{equation}
\label{eq_dV_meas}
\overline{\Delta g} = T^{-2} \left(
\int_T^{2T} \Delta V(t) \D t  -
\int_0^{T} \Delta V(t) \D t
\right),
\end{equation}
\begin{eqnarray}
\label{eq_dz_meas}
\overline{\Delta g}
& = T^{-2}\int_0^{2T} \Delta z(t)
\Big(
\delta(t) - 2\delta(t-T) + \delta(t-2T)
\Big) \D t \nonumber \\
 & = \Big(
\Delta z(0) - 2 \Delta z(T) + \Delta z(2T)
\Big)/T^{2}.
\end{eqnarray}
It's interesting to observe how the formulas (\ref{eq_dz_meas}) and (\ref{eq_3 points}) converge, highlighting the fact that the gravimeter's weighting function by coordinate is just the finite Dirac comb providing sampling of the continuously changing coordinate.
\section{Effective measurement height in ballistic determinations of gravity}
The constant vertical gradient $\gamma$ changes gravity with height like
\begin{equation}
\label{eq_g_z}
g(z) = g_0 + \gamma z,
\end{equation}
where $g_0$ is gravity at some initial height, $z$ is the distance from that height. If the measured gravity can be presented as
\begin{equation}
\label{eq_g_z_meas}
\overline g = g_0 + \gamma \; h_{\bf{eff}},
\end{equation}
where $h_{\bf{eff}}$ is some gradient-independent value, then it follows from comparison of (\ref{eq_g_z}) and (\ref{eq_g_z_meas}) that
\begin{equation}
\label{eq_g_z_g_meas}
\overline g = g(h_{\bf{eff}}),
\end{equation}
which means that $h_{\bf{eff}}$ represents a distance from the initial position to the point at which the measured gravity equals local gravity. The locus of that point is called the effective measurement height of the instrument. The equation (\ref{eq_g_z_g_meas}) holds true for any constant gradient. Reporting the measured gravity at the effective measurement height does not involve the gradient value, and so its uncertainty does not affect the total uncertainty of the measurement. If the measured gravity is reported at the initial position, the uncertainty of the gradient $\Delta \gamma$ propagates into the uncertainty of the measurement as
\begin{equation}
\label{eq_g_grad_uncert}
\Delta \overline g = \Delta \gamma \; h_{\bf{eff}}.
\end{equation}
\section{Effective measurement height of atom gravimeters}
\subsection{Release gravimeters}
In this type of instruments the atoms are released for the free fall at some initial height. The measurement starts at the moment $t_1$ when the atoms' separation from the initial position is $z_1$.
Due to the vertical gravity gradient $\gamma$ the gravity changes in time like
\begin{equation}
\label{eq_dg_grad}
g(t) = g_0 + \Delta g_{\gamma}(t) = g_0 + \gamma z(t) =
g_0 + \gamma (z_1 + V_0 t + g_0 t^2/2).
\end{equation}
We associate the initial velocity $V_0$ with the moment $t_1$, at which the measurement starts, while $g_0$ is related to the initial position of the atoms. The changing acceleration (\ref{eq_dg_grad}) translates, according to (\ref{eq_dg_meas},) into the following measured gravity
\begin{eqnarray}
\label{eq_delta_g_grad_meas}
\overline g & = g_0 +
\gamma \int_0^{2T} (z_1 + V_0 t + g_0 t^2/2) \; w_g(t) \D t
 \nonumber \\
& = g_0 + \gamma(z_1 + V_0 T + \frac{7}{12} g_0 T^2) = \gamma \; h_{\bf{eff}}.
\end{eqnarray}
Let $h$ be the length of idle part of the trajectory, $H$ be the total trajectory length. Obviously, $h=z_1$, $H= z_3$. Taking into account that $V_0 = g_0 z_1$, $T=(t_3 - t_1)/2 $, $t_1^2 = 2z_1/g_0$, $t_3^2 = 2z_3/g_0$, we obtain
\begin{equation}
\label{eq_h_z0}
h_{\bf{eff}} = \frac{7}{24}(H+h) + \frac{5}{12}\sqrt{Hh}.
\end{equation}
\subsection{Fountain gravimeters}
In this type of instruments the atoms are thrown up vertically, the gravity is measured on both upward and downward parts of the trajectory. Let's direct $z$-axis vertically and consider an atom thrown up along the axis. Like before, we associate the atom's initial velocity $V_0$ with the start of the measurement at  moment $t_1$. Due to the gradient, the gravity changes like
\begin{equation}
\label{eq_g_upward}
g(t) = g_0 - \gamma(z_1 + V_0 t - g_0 t^2/2).
\end{equation}
The measured gravity, according to (\ref{eq_dg_meas}), is
\begin{eqnarray}
\label{eq_g_upward_meas}
\overline g & =  g_0 - \int_0^{2T}\gamma(z_1 + V_0 t - g_0 t^2/2)w_g(t) \D t \nonumber \\ & =
g_0 + \gamma
\left(\frac{7}{12}g_0 T^2 - V_0 T- z_1\right).
\end{eqnarray}
The result obtained in (\ref{eq_g_upward_meas}) using the LTI approach coincides with the results obtained with both the path integral and the perturbation Lagrangian methods \cite{peters2001}. Let's now find the effective measurement height of the fountain gravimeter. In order for an atom to reach the apex of the trajectory in time $T$, the magnitude of the $V_0$ should be $g_0T$.
Substituting this value into (\ref{eq_g_upward_meas}) we get
\begin{equation}
\label{eq_g_upward_meas2}
\overline g =
g_0 + \gamma
\left(\frac{7}{12}g_0 T^2
- g_0 T^2- z_1
\right)
= g_0 - \gamma
\left( \frac{5}{12}g_0 T^2 + z_1 \right)
.
\end{equation}
As in fountain gravimeters the total height is $H=gT^2/2$, we get
\begin{equation}
\label{eq_g_sym_wo_Vo}
\overline g =
g_0 - \gamma
\left( \frac56 H + z_1 \right),
\end{equation}
which means that the effective measurement height for the fountain gravimeters is located above the initial position at
\begin{equation}
\label{eq_g_heff_fountain}
h_{\bf{eff}}= \frac56 H + z_1.
\end{equation}
As the gravity at the apex of the trajectory is $g_0 - \gamma(H+z_1)$, we can express the same as the following distance below the apex:
\begin{equation}
\label{eq_g_sym_heff}
h'_{\bf{eff}}=\frac16 H.
\end{equation}
\section{Conclusions}
We have found the position of the point corresponding to the effective measurement height of the release and the fountain types of atom gravimeters. At that point the local gravity of the field with the constant vertical gradient is the same as the measurement result of atom gravimeter when it's not corrected for the gradient. Leaving the gravity value for reporting at the effective measurement height allows to eliminate the error associated with the uncertainty of the gradient. For many gravity sites this uncertainty can not be estimated better than 10 $\mu$Gal m${^{-1}}$ \cite{csapo2004}. For the 12 cm of total trajectory length of release gravimeters \cite{bodart2010} this uncertainty, according to (\ref{eq_h_z0}) and (\ref{eq_g_heff_fountain}), translates into additional 0.35 $\mu$Gal of uncertainty of the measured gravity value. For about the same trajectory length of the fountain gravimeters \cite{peters2001}, the additional uncertainty of the measured gravity, according to (\ref{eq_g_sym_wo_Vo}) and (\ref{eq_g_heff_fountain}), can reach 1 $\mu$Gal.
\section*{References}
%
%\bibliography{disser}

\bibliographystyle{ieeetr}
\end{document}